\documentclass[sigconf]{acmart}

\usepackage{booktabs} % For formal tables

\copyrightyear{2017}
\acmYear{2017}
\setcopyright{acmcopyright}
\acmConference{GECCO '17}{July 15-19, 2017}{Berlin, Germany}\acmPrice{15.00}\acmDOI{http://dx.doi.org/10.1145/3071178.3071336}
\acmISBN{978-1-4503-4920-8/17/07}

\begin{document}
\title[Behaviorally-Functional Networks of Spiking Neurons]{Evolution and Analysis of Embodied Spiking Neural Networks Reveals Task-Specific Clusters of Effective Networks}
%\titlenote{Produces the permission block, and copyright information}
%\subtitle{Extended Abstract}
%\subtitlenote{The full version of the author's guide is available as
%  \texttt{acmart.pdf} document}

\author{Madhavun Candadai Vasu}
\affiliation{%
  \institution{Indiana University}
  %\streetaddress{P.O. Box 1212}
  \city{Bloomington}
  \state{Indiana}
  \postcode{47405}
}
\email{madcanda@indiana.edu}

\author{Eduardo J. Izquierdo}
\affiliation{%
  \institution{Indiana University}
  %\streetaddress{P.O. Box 1212}
  \city{Bloomington}
  \state{Indiana}
  \postcode{47405}
}
\email{edizquie@indiana.edu}

\renewcommand{\shortauthors}{Candadai Vasu et. al.}

\begin{abstract}
Elucidating principles that underlie computation in neural networks is currently a major research topic of interest in neuroscience. Transfer Entropy (TE) is increasingly used as a tool to bridge the gap between network structure, function, and behavior in fMRI studies. Computational models allow us to bridge the gap even further by directly associating individual neuron activity with behavior. However, most computational models that have analyzed embodied behaviors have employed non-spiking neurons. On the other hand, computational models that employ spiking neural networks tend to be restricted to disembodied tasks. We show for the first time the artificial evolution and TE-analysis of embodied spiking neural networks to perform a cognitively-interesting behavior. Specifically, we evolved an agent controlled by an Izhikevich neural network to perform a visual categorization task. The smallest networks capable of performing the task were found by repeating evolutionary runs with different network sizes. Informational analysis of the best solution revealed task-specific TE-network clusters, suggesting that within-task homogeneity and across-task heterogeneity were key to behavioral success. Moreover, analysis of the ensemble of solutions revealed that task-specificity of TE-network clusters correlated with fitness.  This provides an empirically testable hypothesis that links network structure to behavior.
\end{abstract}


% We no longer use \terms command


\maketitle

\section{Introduction}

%[Broad motivation. What is the context for this work? Why is it important?]}
There is an interest like never before to understand how nervous systems produce behavior. Although many aspects of the environment and behavior are largely described with continuous variables, including for example the continuous motion of our bodies and other objects in the environment, the neurons that process this information in many organisms operate through discrete-like action potentials. One of the grand challenges of neuroscience is to understand how this continuous stream of information from the environment is represented and processed by spikes and converted back into fluid behaviors.

%[Current empirical approach and its shortcomings: Information theory in empirical studies and the need for computational modeling]}
One of the most useful tools has been the ability to quantify information transfer by action potentials through the use of information theory~\cite{Rieke1996}.
Transfer entropy (TE)~\cite{Schreiber2000,Kaiser2002} is an information-theoretic measure that is being used extensively for estimating effective networks that emerge from task-specific interactions between the underlying structural network elements at different spatial and temporal scales~\cite{Vincente2011,Ito2011,Rubinov2010,Nigam2016,Gourévitch2007,Garofalo2009,Timme2014,Shimono2015}.
Most of the work analyzing information processing in the brain has been empirical, focusing on three main techniques: fMRI, {\em in vitro}, and {\em in vivo} studies.
While fMRI studies have yielded lots of insights into whole-brain organization, development and pathology, network nodes are defined at the macro or meso-scale where each node corresponds to hundreds or thousands of neurons~\cite{Vincente2011,Kriegeskorte2006}.
{\em In vitro} studies have helped understand the micro-level structural organization and flow of information across small networks of neurons, but it is not possible to study neural dynamics in the context of behavior~\cite{Timme2014, Ito2011}.
Finally, {\em in vivo} studies make it possible to record micro-scale activity from behaving animals, but resolving the cortical network that is involved in a behavior and recording from all neurons involved in that particular behavior is not feasible yet~\cite{Nigam2016}.

% [Current computational approach and its shortcomings: Embodiment]}
% For this paragraph to work, the importance of spiking neurons should have been stated in the first paragraph.
Over the past few decades, theoretical neuroscientists have begun to address these issues by studying computational models of networks of spiking model neurons. % without substantial connections to function?
However, most of this work has focused on characterizing the dynamics of the abstract network, without any substantial connections to function~\cite{Vreeswijk1996,Kopell2000,Brunel2000,Schaffer2013}.
More recently, work has begun to focus on developing functional spiking neural networks~\cite{Abbott2016}. However, this has been generated largely for disembodied networks: a network receives a time-series input, and its task is to generate a specified output~\cite{Thalmeier2016}. These models address how specific patterns of neural activity are generated by sensory stimuli or as part of motor actions, but lack the continuous closed-loop interaction between the brain, body and an environment. Incorporating these components to address how a neural circuit works, requires us to develop behaviorally-functional spiking networks. The view that the body and the environment play a crucial role in the understanding of behavior is increasingly accepted~\cite{Chiel1997,Izquierdo2016}. In parallel, there has been some work that has focused on understanding behavior through the development and analysis of whole brain-body-environment models~\cite{Beer2003}. However, with only a few exceptions~\cite{Harvey2005,Floreano2001}, the majority of this work has used non-spiking neural models.

% [What is our approach? How is it different?]
The work presented here takes an entirely different approach to the study of behaviorally-functional spiking neural networks.
We use an evolutionary algorithm to evolve spiking neural controllers for simulated agents performing a visual categorization task.
There are numerous benefits of using this kind of approach to study the neural basis of behavior.
(a) Brain-body-environment models make it possible to relate neural activity to behavior, because unlike empirical studies, all variables of the system are easily accessible. This allows us to take seriously the view that cognition is situated, embodied, and dynamical~\cite{Pfeifer2007,Chiel1997,Varela1991}.
(b) By designing tasks that are deliberately minimal and yet of interest to cognitive scientists~\cite{Beer1996}, we can begin to address questions about information processing in the brain that are relevant to understanding cognition. %It can be described as a task that can be performed with a small network of neurons while still being deemed cognitively-interesting.
(c) The use of a spiking neuron model, which can replicate the dynamics of Hodgkin--Huxley--type neurons with the computational efficiency of integrate-and-fire neurons~\cite{Izhikevich2004}, allow us to use the same tools used by neuroscientists to analyze the model (e.g., Transfer Entropy). %-- Biologically-plausible in order to match biological computational complexity~\cite{Maass1997} and to the extent that it replicates the dynamics of biological neurons~cite{Izhikevich2004}.
(d) By evolving agents, instead of hand-designing them, we are able to make minimal prior assumptions about how various behaviors must be implemented in the circuit. This maximizes the potential to reveal counter-intuitive solutions to the production of behavior. The evolutionary algorithm was used to determine the values of the electrophysiological parameters that optimize a behavioral performance measure. Further, the state-of-the-art in supervised learning of spiking neural networks is not capable of optimizing the kind of embodied models we are building~\cite{Lin2016}.
(e) Typically, each successful evolutionary search produces a distinct set of parameter values, leading to an ensemble of successful models produced over several runs. The properties of this ensemble can be analyzed to identify multiple ways of solving the same problem and to identify recurring principles that underlie the production of behavior.

% [What do we aim to show in this paper?]
This paper has two broad primary aims.
First, to develop a behaviorally-functional spiking neural network for a cognitively-interesting behavior.
Second, to analyze the ensemble of solutions using transfer entropy and derive robust patterns that underlie different approaches to performing the behavior.
We focus on a visual categorization behavior used in previous computer simulation studies~\cite{Beer2003}.
Categorical perception involves partitioning the sensed world through action~\cite{Harnad1987}. That is, the continuous signals received by the sense organs are sorted into discrete categories, whose members resemble one another more than they resemble members of other categories.
The behavior involves two tasks: catching circle-shaped objects and avoiding line-shaped objects.
In specific, we are interested in the following questions:
(1) Does TE form an explanatory bridge between structure and function? That is, does the inferred effective network tell us something that cannot be observed from the structural network alone about how behavior is produced?
(2) Do different aspects of the behavior (e.g., catching some objects versus avoiding others) have their own specialized effective networks? In other words, are the effective networks for minor variations of one task (say, circle catching) more similar to each other than the effective networks of minor variations of the other task (say, line avoiding)?
%And are these robust across different trials of the task with minor variations?
(3) Is having task-specific effective networks indicative of the agent's performance?  More generally, can the degree of task-specificity of the effective networks predict behavioral performance?

% [Paper organization]
The rest of this paper is organized as follows. In the next section, we describe the agent, neural network model, evolutionary algorithm, and transfer entropy analysis used here. In the following section, we discuss the results of the evolutionary simulations, first by examining a particular case and then generalizing to an analysis of the ensemble of evolutionary runs. Finally, we discuss our results in light of the work on multifunctional networks and on focusing empirical experimental design, and we outline ongoing and future work.

\section{Methods}

This section outlines the technical specifications of the agent, environment, and task; the neural controller; the evolutionary algorithm; and the information-theoretic tools that were used in analysis. The entire simulation was programmed in C++ and analyses were carried out using MATLAB and Python.

\subsection{Agent, Environment, and Task}

In previous work~\cite{Beer1996,Beer2003}, model agents were evolved that could ``visually'' discriminate between objects of different shapes, catching some while avoiding others. These experiments were designed to produce evolved examples of categorical perception~\cite{Beer2003}. All details of the agent, environment, and task have been adapted from these previous studies.

The agent has a circular body with a diameter of 30, and can move horizontally as objects fall from above (Figure~\ref{figSetup}A). The agent's  horizontal velocity is proportional to the sum of opposing forces produced by two motors. The agent also has an ``eye'' which consists of 7 vision rays evenly distributed over an angle of $\pi$/6. These rays extend out from the agent's body with a maximum range of 220. If an object intersects a ray within this range, an external input is fed to a corresponding sensory neuron. The value of the input is inversely proportional to the distance at which the intersection occurs, normalized from 0 to 10.

There are two kinds of objects in the world: circular objects and line objects. Circular objects have a diameter of 30 and line objects have length 30.  These objects fall from a height of 275 at some initial horizontal offset with respect to the agent. Objects fall with a constant vertical velocity of -3 and no horizontal motion.

\subsection{Neural Controller}

\begin{figure}
\centering
\includegraphics[width=1\linewidth]{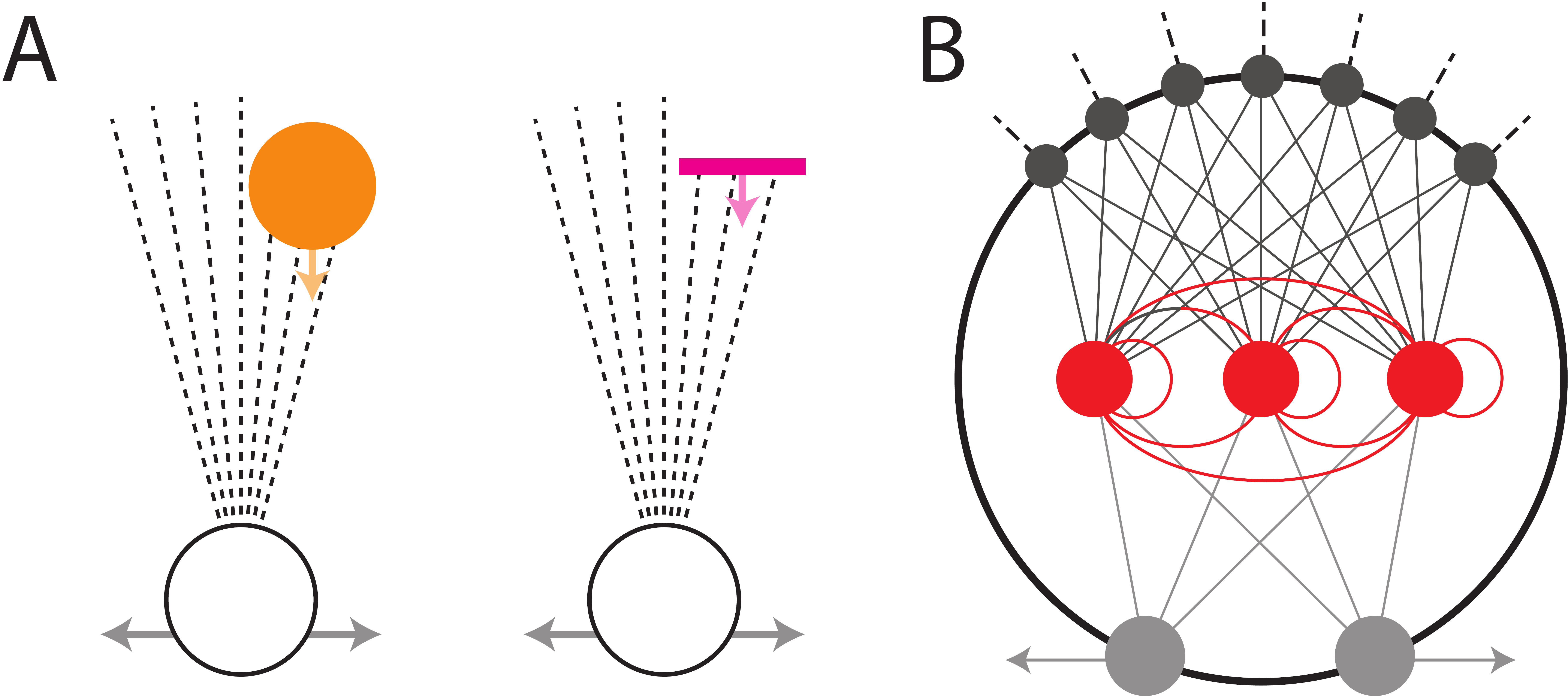}
\caption{Basic setup for the categorical perception experiments. [A]~Agent, environment, and task. The agent moves
horizontally while an object falls towards it from above. The object can be one of two shapes: circle or line. The task consists of catching circles and avoiding lines, adapted from~\cite{Beer2003}. The agent's sensory apparatus consists of an array of seven distance sensors (dashed lines).  [B]~Neural architecture. The distance sensors (black) fully project to a layer of fully interconnected spiking interneurons (red), which in turn fully project to the two motor neurons (light gray).}
\label{figSetup}
\end{figure}

The agent's behavior is controlled by a 3-layer neural network (Figure~\ref{figSetup}B). The network architecture consists of seven sensory neurons fully connected to $N$ fully interconnected interneurons, which are in turn fully connected to two motor neurons.

There are 7 sensory neurons in the top layer which are stimulated by the agent's vision ray. They follow the state equation:

\begin{equation} \label{eq:interneuron}
\tau_s \dot{s_i} = - s_i + K_i(x,y) \quad
i = 1,...,7
\end{equation}

\noindent where $s_i$ is the state of sensory neuron $i$, $\tau_s$ is the time constant that is shares across all sensory neurons, $K_i(x,y)$ is the sensory input from the $i^{\text{th}}$ ray due to an object at location $(x, y)$ in agent-centered coordinates, and the dot notation over the state variable indicates the time differential $\frac{d}{dt}$.

The seven sensory neurons project down to a middle layer of $N$ fully interconnected Izhikevich spiking neurons with the following two-dimensional system of ordinary differential equations~\cite{Izhikevich2003}:

\begin{equation} \label{eq:interneuron}
\dot{v_i} = 0.04v_i^2 + 5v_i + 140 - u_i + S_i + I_i \quad
i = 1,...,N
\end{equation}

\begin{equation} \label{eq:interneuron}
\dot{u_i} = a(bv_i - u_i)
\end{equation}

\noindent with the auxiliary after-spike resetting

\[
   \text{if} \; v \geq 30 \text{mV,} \quad \text{then} \;
\begin{cases}
    v \gets c\\
    u \gets u+d
\end{cases}
\]

\noindent with each interneuron receiving weighted input from each sensory neuron:

\begin{equation} \label{eq:interneuron}
S_i = \sum_{j=1}^7 w_{ji}^s \sigma(s_j + \theta_s)
\end{equation}

\noindent and from other spiking interneurons:

\begin{equation} \label{eq:interneuron}
I_i = \sum_{j=1}^N w_{ji}^i o_i
\end{equation}

\noindent where $v_i$ is representative of the membrane potential of spiking neuron $i$, $u_i$ represents its membrane recovery variable, $w_{ji}^s$ is the strength of the connection from the $j^{\text{\tiny th}}$ sensory neuron to the $i^{\text{\tiny th}}$ spiking interneuron, $\theta_s$ is a bias term shared by all sensory neurons, $\sigma(x)=1/(1+e^{-x})$ is the standard logistic activation function, $w_{ji}^i$ is the strength of the recurrent connections from the $j^{\text{\tiny th}}$ to the $i^{\text{\tiny th}}$ spiking neuron, and $o_i$ is the output of the neuron: 1 if $v_i\geq30$mV, and 0 otherwise. The sign of all outgoing connections from an interneuron depend on its excitatory or inhibitory nature, as identified by a binary parameter. Parameters $a$,$b$,$c$ and $d$ control the type of spiking dynamics.
%The ranges of these parameters are dependent on the inhibitory or excitatory nature of the neuron (see Fig. 2 from~\cite{Izhikevich2003} for more detail).
For inhibitory neurons $a\in[0.02,0.1]$,$b\in[0.2,0.25]$,$c=-65$ and $d=2$, whereas for excitatory neurons $a=0.02$, $b=0.2$, $c\in[-65,-50]$ and $d\in[2,8]$.

Finally, the layer of interneurons feeds into the two motor neurons, with the following state equation:

\begin{equation} \label{eq:interneuron}
\tau_m \dot{m_i} = - m_i + \sum_{j=1}^N w_{ji}^m \bar{o}_j \quad
i = 1,2
\end{equation}

\begin{equation} \label{eq:interneuron}
\bar{o}_j(t)= \frac{1}{h_j} \sum_{k=0}^{h_j} o_j(t-k)
\end{equation}

%How spiking activity is read out by motor neurons \\ %How motor neuron activations are converted to velocity}
\noindent where $m_i$ represents the motor neurons, $w_{ji}^m$ is the strength of the connection from the $j^{\text{\tiny th}}$ spiking interneuron to the $i^{\text{\tiny th}}$ motor neuron, $\bar{o}_j$ represents the moving average over a window of length $h_j$ for the output of spiking interneuron $j$.

Finally, the difference in output between the motor neurons results in an instantaneous horizontal velocity that moves the agent in one direction or the other.

Altogether, a network with $N$ interneurons has a total of $P=N^2+14N+4$ parameters. Unlike previous work~\cite{Beer1996,Beer2003}, bilateral symmetry in network parameters was not enforced. States were initialized to 0 and circuits were integrated using the forward Euler method with an integration step size of 0.1.

% DECIDE LATER: IF REQUIRED
%\begin{figure}
%\centering
%\includegraphics[width=\linewidth]{figures/fig1}
%\caption{Different spiking dynamics of an Izhikevich neuron %(Izhikevich, 2003).}
%\label{fig1}
%\end{figure}

\subsection{Evolutionary Algorithm}

A real-valued genetic algorithm was used to evolve the controller parameters: connection weights, biases, time constants, and intrinsic neuron parameters. Agents were encoded as $P$-dimensional vectors of real numbers varying from [-1, 1]. Each vector element linearly mapped to a parameter of the circuit: interneuron and motor neuron biases $\in [-4, 4]$, sensory neuron biases $\in [-4, -2]$, time-constants ranged $\in [1, 2]$, and connection weights $\in [-50, 50]$. As mentioned previously, the range of the intrinsic parameters of the Izhikevich neurons and the polarity of their weights depend on a binary parameter that decides the inhibitory/excitatory nature of the neuron. Parents were selected with a rank based mechanism, with an enforced elitist fraction of 0.04. Offspring were generated from uniform crossover of two parents (probability 0.5). A Gaussian distributed mutation vector was applied to each parent ($\mu=0$, $\sigma^2=0.5$).

Fitness was calculated across 48 trials with objects dropped uniformly distributed over the range of horizontal offsets [-50,+50]. The performance measure to be maximized was: $\sum_{j=1}^{48} p_i/48$, where $p_i=1-|d_i|$ for the objects that need to be caught and $p_i=|d_i|$ for the objects that need to be avoided, and $d_i$ is the horizontal distance between the centers of the object and the agent when their vertical separation goes to zero on the $i$th trial. The distance $d_i$ is clipped to a maximum value of 45 and normalized to run between 0 and 1.

\subsection{Transfer Entropy}
Transfer entropy (TE) is an information-theoretic measure which has received recent attention in neuroscience for its potential to identify effective connectivity between neurons~\cite{Wibral2014}. Intuitively, TE from neuron $J$ to $I$ is a measure of the additional information provided by the activity in neuron $J$ over and above the information from $I$'s own history of activity that helps predict the activity of neuron $I$. This proportional increase, when high, corresponds to a causal influence of $J$ over $I$. TE is especially useful in taking into account non-linear interactions between neural units and provides a directed measure of influence from one neuron to another. Further, due to synaptic delays, the causal influence of one neuron over another can only be detected if tested at that corresponding delay. In order to account for this, a modified version of TE was proposed by Ito et. al~\cite{Ito2011}. We used the MATLAB toolbox provided by the these authors to estimate TE. This involved utilizing the history of neuron $J$ over different time delays to predict the future activity of $I$ and then picking the peak-TE over all delays, as follows:

\begin{equation} \label{eq:interneuron}
\text{TE}_{J\to I}(d) = \sum p(i_t,i_{t-1},j{t-d}) \text{log}_2 \frac{p(i_t|i_{t-1},j_{t-d})}{p(i_t|i_{t-1})}
\end{equation}

\noindent where $i_t$ denotes a binary spike/no-spike activity of neuron $I$ at time $t$, where $j_t$ denotes a binary spike/no-spike activity of neuron $J$ at time $t$, $d$ denotes the synaptic time delay, $p(x)$ is the probability of that set of spiking events occurring at that particular times, and $p(x|y)$ is the conditional probability that a set of spiking events occur given that certain other events have occurred.

\subsection{Cluster Specialization Coefficient}

Successful agents were subject to TE analyses on a trial-by-trial basis. That is, the trial-specific TE network was estimated for one trial of falling line or circle. This produces 48 TE networks from the 24 circle and 24 line trials. These networks are then hierarchically clustered using a vanilla agglomerative clustering algorithm in MATLAB. This procedure produces a clustering tree diagram where the leaves of the tree correspond to one of the 48 networks. The leaves are successively connected to one another depending on how close they are in TE network space, until they are all in one cluster. These trees are called dendrograms. For each successful agent we produced a trial-by-trial TE dendrogram.

In order to quantify the task-specificity of the TE networks, we measured the number of clusters that were unique to each task.  We cut the dendrogram at different places to partition the 48 networks into multiple clusters. Then the members of the cluster were sorted based on task. We identified the smallest number of clusters that can be formed, such that all the networks in the cluster correspond to networks that were inferred from the neural activity of one and only one of the tasks. These clusters are called task-specialized clusters.
For example, the dendrogram in Figure~\ref{figTEBN}C can be dissected at level 1, to partition the data into two task-specialized clusters, each containing only networks corresponding to one task.
Further dissection of these clusters will yield more number of task-specialized clusters but the minimum number of task-specialized that can be formed is 2.
CSC is defined based on the ratio between the minimum number of task-specialized clusters to the maximum number of task-specialized clusters that can be formed.
The maximum number of task-specialized clusters that can be formed is equal to the number of networks i.e. one network per cluster. Therefore, the CSC for an agent is defined as
$\text{CSC} = 1 - c_{\text{Min}}/c_{\text{Max}}$, where $c_{\text{Min}}$ is the minimum number of task-specialized clusters and $c_{\text{Max}}$ is the maximum number of task-specialized clusters.

%\begin{equation}
%\text{CSC(i)} = 1 - \frac{\text{min number of %task-specialized clusters}}{\text{max number of %task-specialized clusters}}
%\end{equation}

\section{Results}

\subsection{Minimal spiking circuit for object categorization task}

\begin{figure}
\centering
\includegraphics[width=1\linewidth]{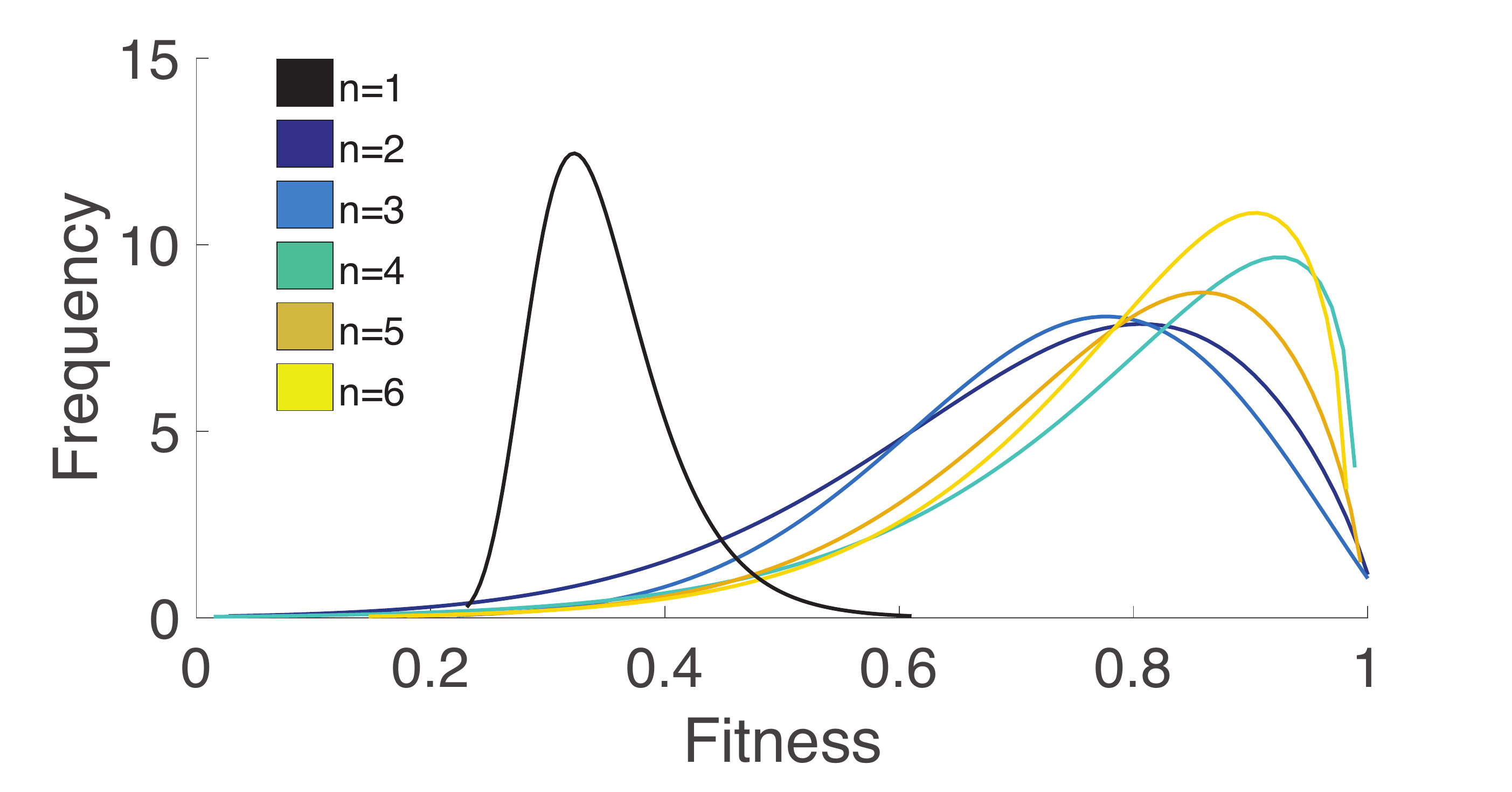}
\caption{Fitness distributions for the best evolved agents from 100 evolutionary runs for network sizes 1-6. No network with only one spiking neuron evolved to solve the task. Networks containing as little as two spiking neurons were found that could solve the task. Interestingly, larger networks worked just as well as networks of size two.}
\label{fig_FitDist}
\end{figure}

In order to identify the smallest network that could perform the  visual categorization task, we ran 100 evolutionary runs with different spiking network sizes ranging from 1 through 6.
In Figure~\ref{fig_FitDist} we show the best fitness after 1000 generations for each of the evolutionary runs, for all the circuit sizes, as a smoothed histogram distribution.
Other than the circuits with only one spiking neuron, all evolutionary runs produced circuits that could solve the task.
Interestingly, networks with only two spiking neurons evolved just as reliably as larger networks.

\subsection{Analysis of the best agent}
\begin{figure}
\centering
\includegraphics[width=0.9\linewidth]{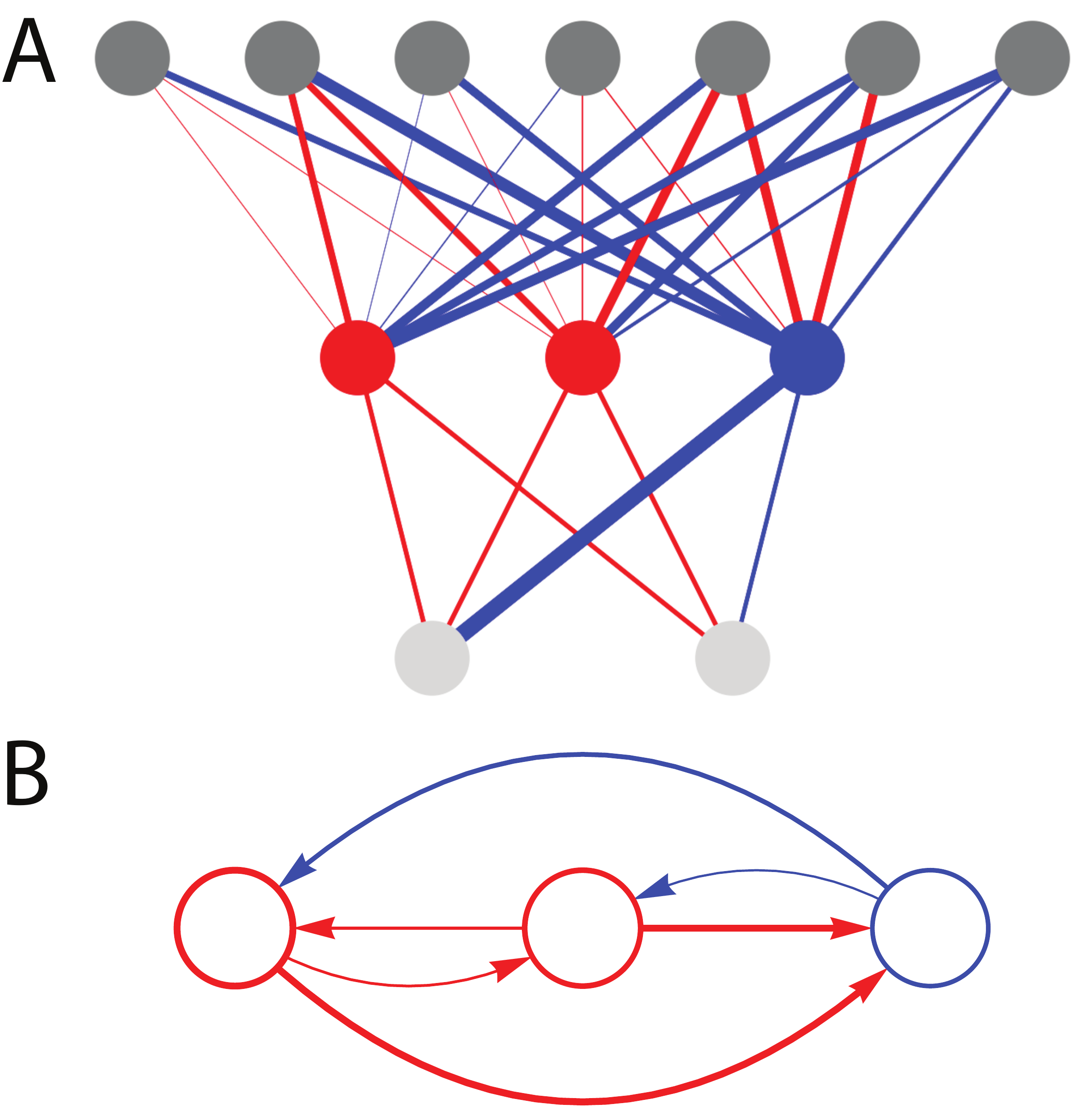}
\caption{Structure of the best $N=3$ network. [A] Feed-forward component. Top layer represents the sensory neurons. Middle layer represents spiking interneurons. Bottom layer represents motor neurons. Excitatory connections shown in blue; inhibitory connections in red. The spiking neurons are also colored according to whether they are excitatory or inhibitory. [B] Interneuron recurrent component. Strength of self-connection is shown by the thickness of the ring on the node. Relative strengths of connections are shown by the thickness of the edges in A and B.}
\label{figStructureBN}
\end{figure}

We selected the highest fitness individual (fitness=97.8\%) over all 100 runs from the network size $N=3$ batch to analyze first. The spiking network for this individual was composed of 2 inhibitory neurons and 1 excitatory neuron (Figure~\ref{figStructureBN}). The feed-forward component shown in Figure~\ref{figStructureBN}A captures the sensory neurons to interneuron weights and the interneuron to motor neuron weights represented by the thickness of the edges connecting the nodes. Figure~\ref{figStructureBN}B shows the fully-connected recurrent inter-layer weights where thickness of the edges denote weights and the width of the nodes denote the self-connection. In both cases, red edges denote inhibitory weights and blue edges denote excitatory weights.

\begin{figure}
\centering
\includegraphics[width=0.95\linewidth]{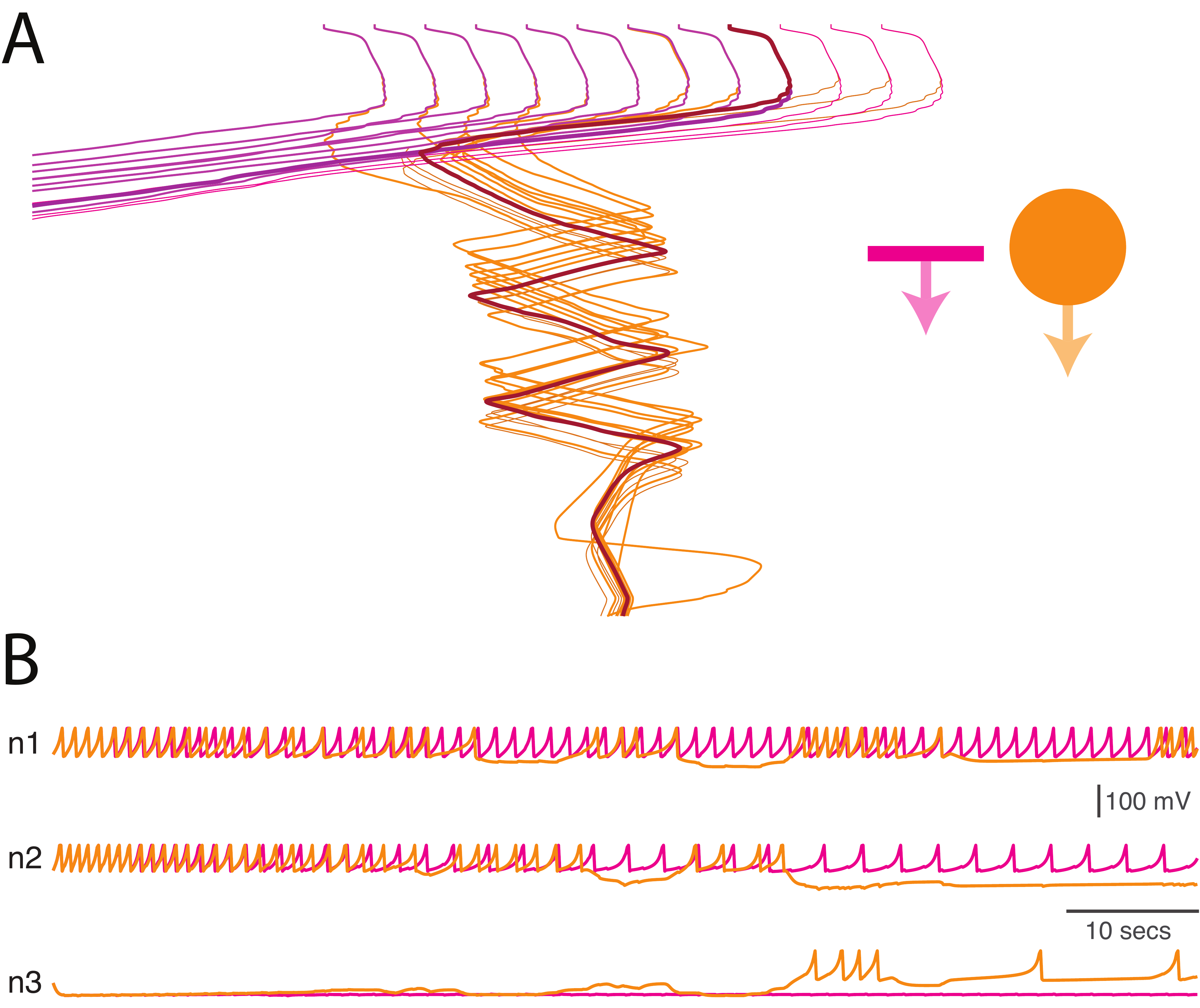}
\caption{Behavioral dynamics of the best network. [A]~Behavior. The horizontal offset between the agent and the falling object is shown along the horizontal access. The vertical axis denotes the time. As the object falls, it can be seen that the orange traces (representing the circles) end up with 0 offset, meaning the the agent ``caught'' them and vice versa for lines. Only 24 of the 48 trials are shown for clarity. Two traces are highlighted. [B]~Neural traces for example runs highlighted in [A]. The neural activity is similar at the beginning of the object's fall. Note that neuron 3 does not show any activity for the line avoiding task but does spike during the circle catching task.}
\label{figBehaviorBN}
\end{figure}

This agent perfectly differentiated the circles and the lines by ``catching'' all circles and moving away from all the lines (Figure~\ref{figBehaviorBN}A). The horizontal offset distance between the agent and falling object as it approaches the agent is 0 for all circles and is large for all lines. The fitness of the agent is not a perfect 100\% simply because that would entail catching the circles at exactly the center of its 30-unit wide body. The small dispersion at the end of the behavioral traces in Figure~\ref{figBehaviorBN}A is representative of the agent's small deviation from catching at its center. Visualizing the corresponding spiking neuron activity (Figure~\ref{figBehaviorBN}B) shows that the inhibitory neurons ($n1$ and $n2$) have a higher firing rate than the excitatory neuron. This is consistent with relative firing rates in biological neurons~\cite{Wilson1994}.

\begin{figure}
\centering
\includegraphics[width=\linewidth]{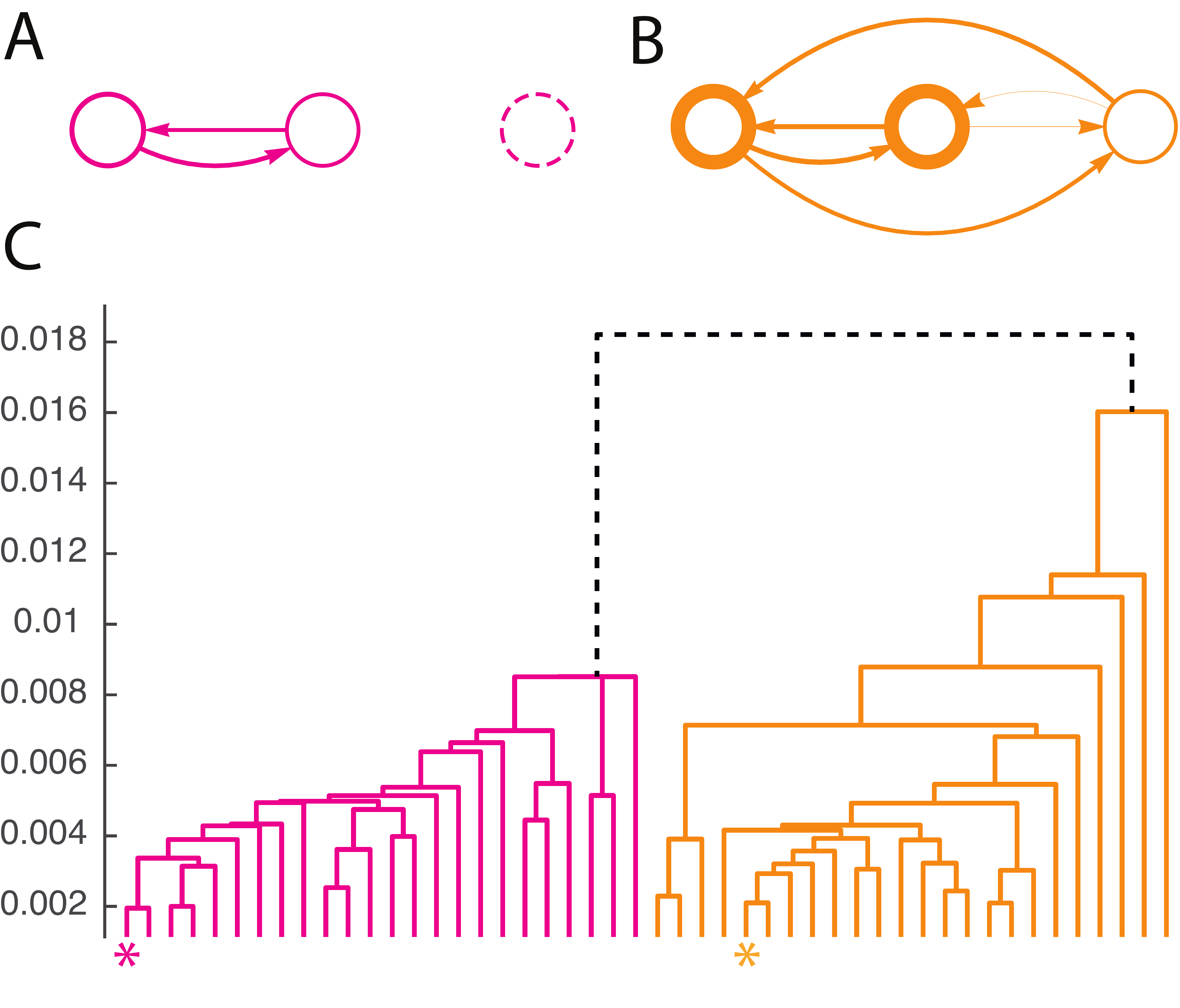}
\caption{Transfer entropy networks of the best $N=3$ agent. [A]~Effective network for one line avoiding trial highlighted in Figure~\ref{figBehaviorBN}A. Note that neuron 3 is not part of this effective network. This is consistent with its lack of activity for this task, as shown in Figure~\ref{figBehaviorBN}B. [B]~Effective network for one circle catching trial highlighted in Figure~\ref{figBehaviorBN}A. [C]~A dendrogram of the hierarchical clustering of trial-by-trial TE networks. The branches of the tree are color-coded by whether the TE network was inferred from a circle catching task (orange) or line avoiding task (magenta). It can be seen here that the tree can be cut at the top level to produce 2 clusters whose members are of only one task. The TE networks shown in panels [A] and [B] are identified in this tree with color matched asterisks.}
\label{figTEBN}
\end{figure}

In order to probe the difference in neural dynamics between the two tasks, the agent's spiking neuron activity from each of the 48 trials of falling lines and circles was recorded and analyzed. We used Transfer Entropy to estimate the task specific effective network for each trial. Since TE quantifies information transfer in the comparable units of bits, we can directly compare TE networks from spiking activity recorded while the agent is performing different tasks. The effective network estimated from one line-avoiding trial (Figure~\ref{figTEBN}A) shows that $n3$ is not part of the network. This can be reconciled with its activity shown in magenta in Figure~\ref{figBehaviorBN}B which shows that $n3$ is completely dormant during this task. The effective network for one circle catching trial (Figure~\ref{figTEBN}B) on the other hand, shows a fully connected network. Spiking activity during one of the circle catching trials (orange traces in Figure~\ref{figBehaviorBN}B) shows that the neural activity in $n1$ and $n2$ control the scanning behavior of the agent, while $n3$'s activity coincides with the agent's fine motor control at the end of the trial when it needs to center itself with the circle. This explains the lack of participation by $n3$ in the line avoiding task because such fine motor skills are not required.

It is to be noted that depending on the activity in the neural network, the estimated TE networks significantly differ from the structural network. Another point to note is that the TE networks also include recurrent self-connections and these are computed simply based on a neuron's ability to predict its own behavior. While TE networks for only one circle and one line trial are shown here, we performed the same analysis for all trials on this best agent. We found that all the observations made for the individual trials were consistent across the rest of the trials.

\subsection{TE network clusters show task specialization} The relationship between different TE networks across all trials for the best $N=3$ agent was determined by clustering of the 48 TE networks. This resulted in the networks getting organized by task. This observation was made by first estimating the task-specific TE networks for each of the 48 trials (24 circle TE networks and 24 line TE networks). These 48 networks were then clustered using a simple hierarchical clustering algorithm. The dendrogram of the clustered TE networks shown in Figure~\ref{figBehaviorBN}C shows an interesting property. All TE networks of the same task (either circles or lines) fall under one cluster before being grouped into the cluster of the other task. This means that the same structural network effectively presents itself as very similar networks for variations of one task, which are all different from the very similar effective networks for the other task. In other words, there is high within-task homogeneity and high across-task heterogeneity in the task-specific effective networks. Naturally, the next step is to look for this phenomenon in other $N=3$ agents.

\subsection{High CSC agents have high fitness} We developed a metric to quantify and compare task-specialization in TE network clusters, which henceforth we will refer to as cluster specialization coefficient, or CSC. $N=3$ agents that had high CSC showed very high behavioral performance. The greater the value of CSC, the greater the expression of within-task homogeneity and across-task heterogeneity, with the maximum value being 0.95 corresponding to the 2 specialized clusters that purely have effective networks corresponding to one and only one task in each of them. All agents, from the 100 evolutionary runs that had a CSC of 0.95 were collected and they were all high-performing agents. It is to be noted that the structural networks of agents that had a high CSC were highly degenerate in terms of the ratio of inhibitory-excitatory neurons and yet exhibited this phenomenon.

\begin{figure}
\centering
\includegraphics[width=0.95\linewidth]{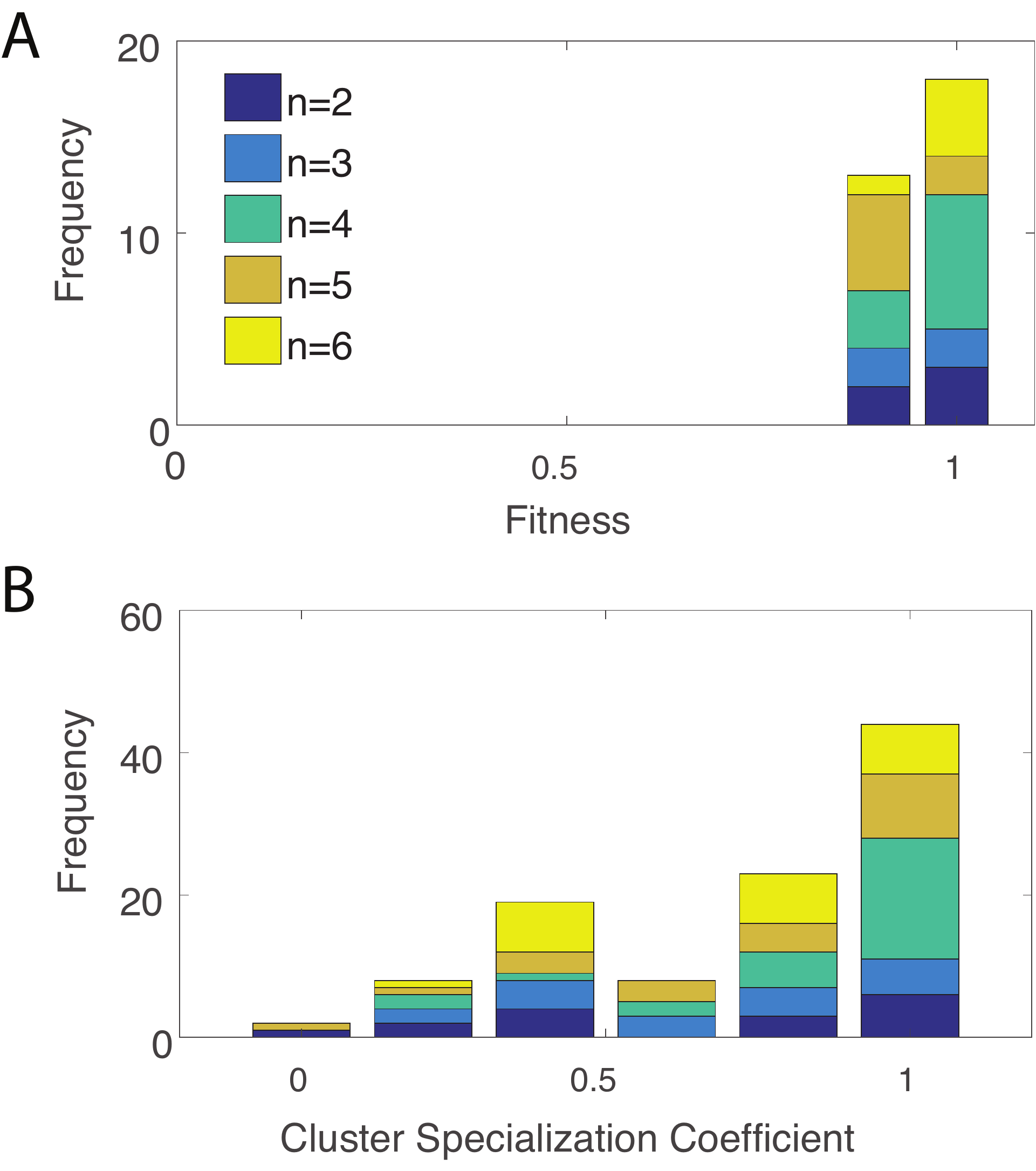}
\caption{Performance and CSC distributions of the high-performing agents for all $N$. [A] A histogram of the fitness of all individuals that have the highest CSC of 0.95 showing that all agents that had a high CSC value performed well. [B] A histogram of the CSC values for agents that evolved to have a fitness $\geq$ 90\%, showing that while there are agents that perform well with low CSC values, it is more likely that an agent with high fitness also has a high CSC value.}
\label{fig4}
\end{figure}

\subsection{High fitness with high CSC is independent of network size} The same analyses that were performed on the best $N=3$ agent and other $N=3$ agents, were repeated for smaller ($N=2$) and larger networks ($N=4$,5 and 6). Irrespective of the network size, agents that have the maximum possible CSC of 0.95, consistently showed high behavioral performance (Figure~\ref{fig4}A). This establishes the idea that high within-task homogeneity and high across-task heterogeneity of task-specific effective networks yields high performance in this task. The obvious next question is to ask if the converse is true. Plotting the distribution of CSCs for agents that have $\geq$90\% fitness  (Figure~\ref{fig4}B) revealed that there exists solutions that have CSCs as low as 0 that still perform well. However, a majority of agents that perform well have a high CSC.

\begin{figure}
\centering
\includegraphics[width=\linewidth]{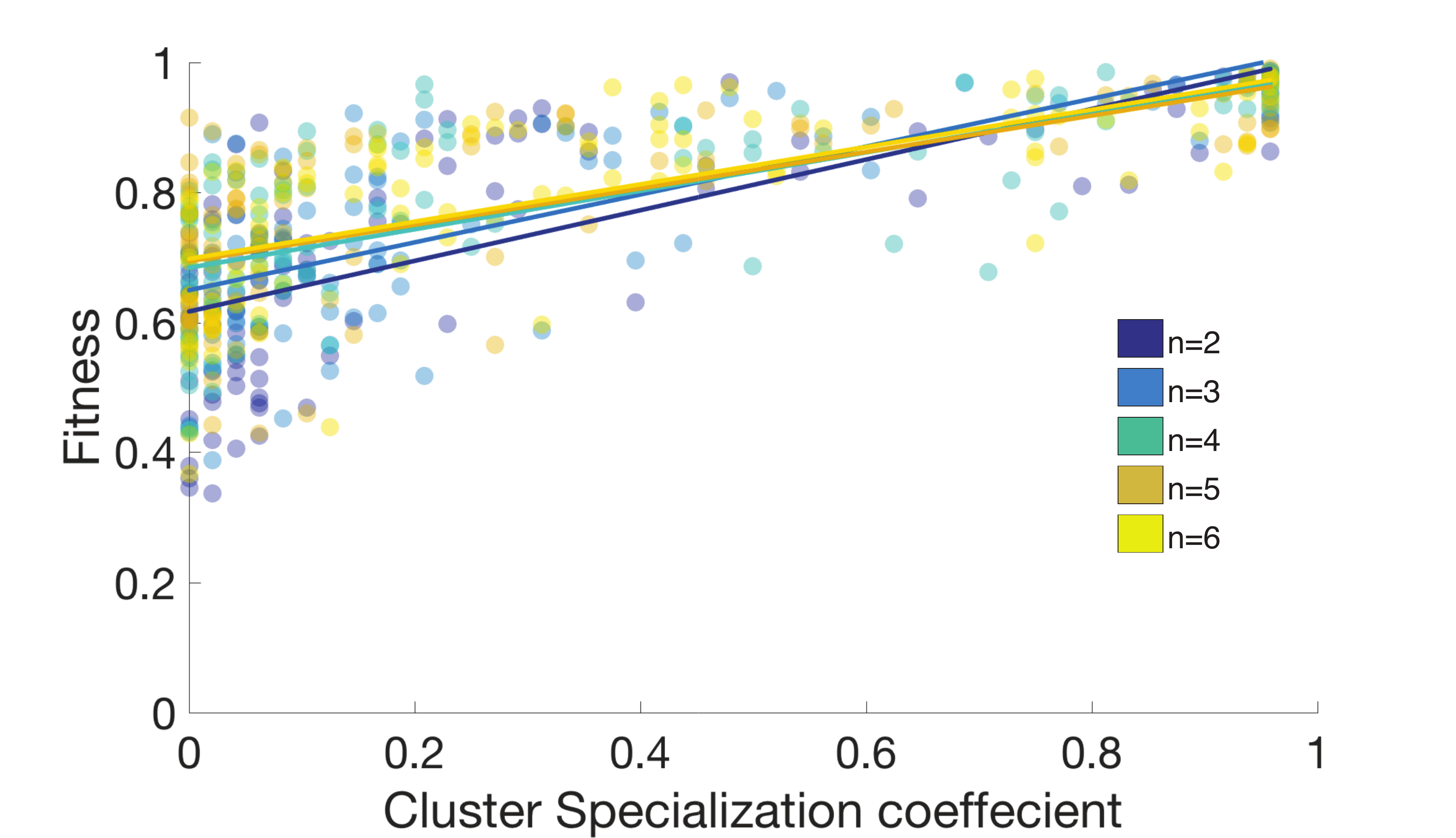}
\caption{High task-specificity in the effective network suggests high performance. Fitness as a function of CSC for all best-agents across the ensemble, across different network sizes. The  high-concentration of points in the right top corner reveals that a high CSC makes it more likely that behavioral performance is also high. Linear regression lines show a positive correlation between fitness and CSC for all network sizes. Interestingly there is no significant difference in this trend for different network sizes.}
\label{fig5}
\end{figure}

\subsection{CSC positively correlates with behavioral performance} In order to further ascertain the relationship between CSC and fitness we looked at the fitness of all agents as a function of their CSC. As shown in Figure~\ref{fig5} fitting a straight line to the points on a CSC versus Fitness plot, shows a positive correlation between an agent's CSC and fitness for all values of $N$. Based on this and the previous result, we conclude that expressing specialized task-specific effective networks is not a necessary condition for high performance but conversely, it is sufficient for high performance. Given the generic nature of the network and it's evolutionary optimization process, it can also be said that this is true for all categorization tasks. It is intuitive that networks that can effectively express themselves distinctly for each category can perform well.

\section{Discussion and Conclusion}
In this paper we present results from the evolutionary optimization of embodied spiking neural networks for cognitively-interesting behavior. We applied information theoretic tools to extract task specific network representations based on spiking dynamics in the interneuron layer. This information showed the interesting characteristic of being clustered by task. We developed a novel metric, Cluster Specialization Coefficient~(CSC), that quantifies the task-specificity of TE networks by dissecting the hierarchical clustering tree of trial-by-trial effective networks. Further analysis of the ensemble revealed a trend that shows positive correlation between CSC and performance. While we also noticed that there are some cases where agents performed well and had a low CSC, we found no agents that performed poorly with a high CSC. Further, we only claim correlation and not causation. However, in combination with neuroscience literature on the existence of task-specific effective networks in the macro scale~\cite{Fox2005}, we hypothesize that biological networks self-organize to have high within-task homogeneity and high across-task heterogeneity for the purposes of efficient categorization of stimuli.

It is widely known that biological neural networks perform multiple functions using the same underlying structural neural circuit~\cite{Blitz1999,Briggman2006}. Our work provides a framework to study the task-specific effective networks that emerge from these otherwise structurally-similar networks. While we have analyzed this system for different tasks in a single behavior, it can be easily extended to multiple-behaviors. The same analysis can be performed to study the task-relevant variations in the neural dynamics and its robustness across variations of the task. Another perspective to look at this from, is that of neural reuse~\cite{Anderson2010}. Presenting new tasks in evolutionary time, would allow the networks to reuse behaviorally appropriate functional components developed for previously evolved behaviors for the new behavior. While it is obvious that the same structural components are reused, functional reuse can be analyzed by comparing the task-specific effective networks. Furthermore, systematic studies of multiple behaviors can further explain the relationship between different types of behaviors and neural reuse.

Our results have revealed a relationship between different categorization behaviors and the allocation of neural resources to perform them: acquiring category-specific effective networks yields high performance.
We intend to further test this by including the CSC as component in the fitness estimation. This will help determine if encouraging this phenomenon makes it more likely to produce multifunctional circuits. Based on our results, we hypothesize that this will in fact be the case.

The model presented here also speaks to the computational ability of spiking neurons in an embodied context. It has been established spiking neurons have greater computational power than perceptron-like neurons~\cite{Maass1997}. Furthermore, principles of morphological computation have shown that using systems that can exploit the continuous interaction between environment and the agent effectively increases the computational abilities of the neural networks~\cite{Pfeifer2009}. In our model, these two concepts are combined because spiking neurons are now involved in a morphological computation scenario. Although it is difficult to quantify the computational power, that only 2 spiking neurons can perform the behavior is indicative of that.

As a measure of effective network connectivity, although Transfer Entropy has been most widely used in neuroscience, it has been used in evolutionary robotics in some instances - synchronization dynamics of neurons and its influence on evolvability and behavior has been studied using the same task described in this paper~\cite{Moioli2012}, and TE has also been used in analysis of neural networks in supervised and unsupervised learning contexts~\cite{Moioli2013}. One of the limitations of this methodology is that TE does not scale elegantly with the size of the networks. While this is not an issue with neuroscientists where TE is used as a post-task analysis method, an online estimation of TE networks during evolution can be computationally expensive. Although small spiking networks, even as small as 2 neurons like we have shown here, have high computational power~\cite{Maass1997}, this can be a problem in large networks for complex tasks. Therefore, in order to use CSC as a component of fitness to more efficiently develop artificial systems, a computationally more efficient metric might be required. This is one of the directions we plan to explore further. Meanwhile, the limits of TE can be pushed to evolving and analyzing increasingly larger networks for a number of different behaviors and also for performing multiple behaviors. These behaviors can be systematically chosen to be (a)~completely independent, (b)~overlapping or (c)~partially overlapping. Studying the relationship between CSC and fitness in each of these cases would provide interesting insights into neural reuse and how neural dynamics shapes behavior. Another limitation of this study is that the TE analysis was performed on a subset of the evolved parameters: the recurrent connections between interneuron spiking neurons; sensor-to-interneuron weights and interneuron-to-motorneuron weights were not included. Constraining the optimization space to only the recurrent connections and setting other weights constant, would make sure that the TE analysis is performed on the system that is entirely-responsible for controlling the differences in behaviors. Finally, TE averages neural dynamics over the entire trial. Although this is useful for comparisons across trials, in order to get an in-depth understanding of how the network produces behavior, unrolling the information analysis over time will be required.

Once again, the potential of computational modeling in helping scientists focus experimental design has been shown here. Specifically, we have shown that evolutionary algorithms can help optimize embodied behaviorally-functional spiking neural networks, and that analysis of the evolved agents can help generate testable insights. The role played by the body and the environment in producing behavior is increasingly acknowledged by neuroscientists~\cite{Izquierdo2016}. Analysis of computational models of embodied agents can thus provide an appropriate platform to develop hypotheses that can then be tested empirically. Looking forward, stimulation and recording experiments {\em in vitro} and calcium imaging experiments in {\it C. elegans} are ideal biological models to test our hypothesis: high within-task homogeneity and high across-task heterogeneity among task specific clusters of effective networks yield high categorization performance.

\begin{acks}
  The authors would like to thank the anonymous referees for their valuable comments and helpful suggestions.
\end{acks}

\bibliographystyle{ACM-Reference-Format}
\bibliography{sigproc}

\end{document}